# Arbitrarily structured quantum emission with a multifunctional metalens

Chi Li[1,2#*], Jaehyuck Jang[3#], Trevon Badloe[4,5#], Tieshan Yang[1], Joohoon Kim[4], Jaekyung Kim[4], Minh Nguyen[1], Stefan A. Maier[2,6], Junsuk Rho[3,4,7,8*], Haoran Ren[2*], Igor Aharonovich[1,9]

1. School of Mathematical and Physical Sciences, University of Technology Sydney, Ultimo, New South Wales 2007, Australia
2. School of Physics and Astronomy, Monash University, Melbourne, Victoria 3800, Australia
3. Department of Chemical Engineering, Pohang University of Science and Technology (POSTECH), Pohang 37673, Republic of Korea
4. Department of Mechanical Engineering, Pohang University of Science and Technology (POSTECH), Pohang 37673, Republic of Korea
5. Graduate School of Artificial Intelligence, Pohang University of Science and Technology (POSTECH), Pohang 37673, Republic of Korea
6. Department of Physics, Imperial College London, London SW7 2AZ, United Kingdom
7. POSCO-POSTECH-RIST Convergence Research Centre for Flat Optics and Metaphotonics, Pohang 37673, Republic of Korea
8. National Institute of Nanomaterials Technology (NINT), Pohang 37673, Republic of Korea
9. ARC Centre of Excellence for Transformative Meta-Optical Systems, University of Technology Sydney, Ultimo, New South Wales 2007, Australia
# These authors contributed equally to this work

Email address: chi.li@uts.edu.au; jsrho@postech.ac.kr; haoran.ren@monash.edu

## Abstract

*Structuring light emission from single-photon emitters (SPEs) in multiple degrees of freedom is of a great importance for quantum information processing towards higher dimensions. However, traditional control of emission from quantum light sources relies on the use of multiple bulky optical elements or nanostructured resonators with limited functionalities, constraining the potential of multi-dimensional tailoring. Here we introduce the use of an ultrathin polarisation-beam-splitting metalens for the arbitrary structuring of quantum emission at room temperature. Owing to the complete and independent polarisation and phase control at the single meta-atom level, the designed metalens enables simultaneous mapping of quantum emission from ultra-bright defects in hexagonal boron nitride and imprinting of an arbitrary wavefront onto orthogonal polarisation states of the sources. The hybrid quantum metalens enables simultaneous manipulation of multiple degrees of freedom of a quantum light source, including directionality, polarisation, and orbital angular momentum. This could unleash the full potential of solid-state SPEs for their use as high-dimensional quantum sources for advanced quantum photonic applications.*

# 1 Introduction

Quantum emission is pivotal to the realisation of photonic quantum technologies, cryptography, communications, sensing, and computing(*1, 2*). Solid-state single photon emitters (SPEs) such as defects in hexagonal boron nitride (hBN) operate at room temperature and are highly desired due to their robustness and brightness(*3-5*) The conventional way to collect photons from SPEs often relies on a high numerical aperture (NA) objective lens or employing micro-structured antennas such as nanopillars(*6*). While the efficiency of photon collection can be high, these tools lack the capability to manipulate quantum emission - particularly in regard to directionality or phase. Consequently, for achieving any desired structuring of the emitted quantum light source, multiple bulky optical elements such as polarisers and phase plates are inevitably required.

The ability to arbitrarily structure light in its multi-dimensions is of great importance for quantum light sources(*7-13*). In this respect, metasurfaces have transformed the landscape of photonic design, propelling major technological advances spanning from optical imaging(*14, 15*), displays(*16*), and holography(*17-21*), to LiDAR(*22*) and molecular sensing(*23*). Recently, the direct integration of nanoscale emitters into nanostructured resonators and metasurfaces have been realised to collect(*24, 25*) and demonstrate basic tailoring of the SPEs emission(*26-29*). These initial demonstrations, although with limited control over quantum emission, constituted the necessity for flat optics to advance the manipulation of quantum emission according to its multiple degrees of freedom.

Here we address this vision, by designing and engineering a multifunctional metalens, capable of tailoring the directionality, polarisation, and orbital angular momentum (OAM) degrees of freedom simultaneously (Fig. 1A). Specifically, we utilise the metalens to demonstrate a multidimensional structuring of quantum emission from SPEs in hBN, operating at room temperature. Since our designed meta-atoms in the metalens can achieve complete and independent polarisation and phase manipulation, it is possible to encode different wavefronts onto orthogonal polarisations of SPEs. Hence, we demonstrate arbitrary shaping of the directionality of quantum emission by adding phase gratings selective to orthogonal polarisations onto the metalens profile, which are represented by different spots in the momentum space of the metalens (Fig. 1B). In addition, we show that different helical wavefronts can be further added onto the metalens profile, leading to the generation of distinctive OAM modes in orthogonal polarisations of SPEs, which are represented by doughnut-shaped spots with different sizes in momentum space (Fig. 1C). The demonstrated arbitrary wavefront shaping of quantum emission in multiple degrees of freedom could unleash the full potential of solid-state SPEs to be used as high-dimensional quantum sources for advanced quantum photonic applications.

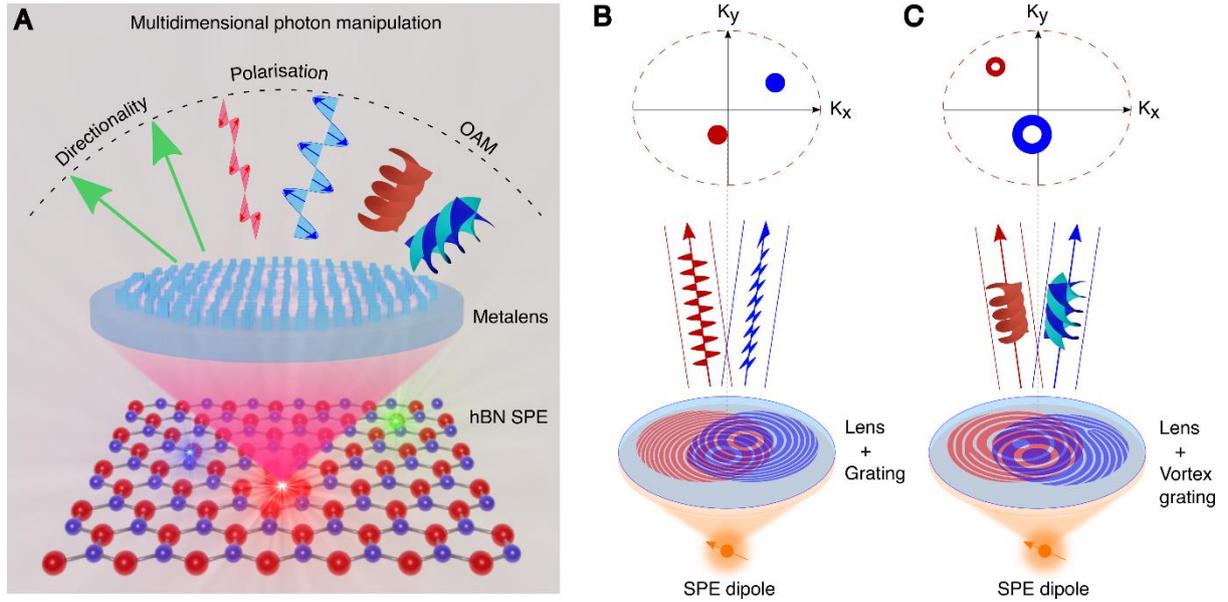

**Fig. 1. Schematics of multidimensional manipulation of hBN quantum emission using a multifunctional metalens.** (**A**) Directional photon splitting, polarisation control and subsequent OAM encoding. The green arrows stand for the directionality control of SPE emission collected by the metalens. The two sine waves indicate the splitting of SPE emission into two orthogonally polarised streams. The two helical beams indicate a further dressing by OAM. (**B, C**) Enlarged views of each encoding concept where grating and vortex grating are adapted to structure the photons in extra dimensions. In addition, the directionality of orthogonal linear polarisations is well inherited and projected as the red and blue spots in the momentum (k) space.

## 2 Results
### 2.1 Design and fabrication of a multifunctional metalens

We design a polarisation beam-splitting metalens to decompose the emission from hBN into two orthogonal linear polarisation (LP) states along the x- and y-axis directions, denoted as LPx and LPy, respectively. We chose the LP to match the linearly polarised emission profile of SPEs in hBN. However, we note that the design could be adapted for other orthogonal polarizations of light, such as left and right circularly polarised light. Taking advantage of the multiplexing possibility offered by the metalens, we add multiple functionalities - specifically controlling the directionality and polarisation, as well as structuring the OAM of quantum emission.

To achieve this, we design cuboid shaped meta-atoms with a fixed height (*H*) and varying length (*l*), and width (*d*), arranged in a square lattice with periodicity (*P*), as shown in Fig. 2A. We choose low-loss hydrogenated amorphous silicon (a-Si:H)(*30, 31*) as the material of the metalens due to its negligible extinction coefficient in the visible emission of hBN SPEs (Supplementary Note 1). We assume that the meta-atoms act as truncated rectangular waveguides, so with a sufficiently large *H*, any desired independent phase modulation ($\varphi$) to the incident LPx and LPy polarisation channels from 0-2$\pi$ can be achieved by varying the transverse dimensions of *l* and *d*. Subsequently, we perform a parameter sweep of different values for *H, l,* and *d,* to design a meta-atom library for matching any required $\varphi$ for LPx and LPy using an in-house-built rigorous coupled wave analysis (RCWA) solver. Eight phase steps between 0-2$\pi$ are successfully achieved from various values of *l* and *d* at a fixed *H* of 550 nm (Fig. 2B). Any meta-atoms with a transmission lower than 80% were ruled out from the library, and the meta-atoms with the closest $\varphi$ for both LPx and LPy were determined for each required phase step. The chosen 8×8 meta-atoms show an average transmission of 94%, and average discrepancy in $\varphi$ of ~0.09 rad. The phase maps of a metalens with a high numerical aperture (NA) of 0.869 were designed to collect and collimate the emission from SPEs. To realise the control of directionality of LPx and LPy emissions, two independent linear phase gratings can be further implemented onto the lens profile of the orthogonal polarisations. The angle of the deflection depends on the grating period that the phase gradient cycles over $2\pi$. As an example, the required phase maps for LPx and LPy to deflect beams in the ±x direction are shown in Fig. 2C, while the whole design process and lens functions are outlined in more detail in Supplementary Note 2.

To fabricate the designed metalens, we employ standard electron-beam lithography followed by a dry etching process (see Methods). Optical microscope as well as scanning electron microscope images are shown in Fig. 2D. The cuboid shaped meta-atoms are fabricated with high accuracy across the metalens, albeit with a slightly slanted sidewall. We experimentally characterise the metalens performance through a dipole-like incidence (Supplementary Note 3). As shown in Fig. 2E, the fabricated metalens can efficiently deflect the orthogonal polarisations of LPx and LPy into different directions, with two distinct collimated beams propagating at around $\pm 16°$. We define our metalens efficiency as the product of collection efficiency corresponding to the maximum collection angle (NA of the metalens) and total transmission efficiency of the metalens (Fig. 2F). The collection efficiency as a function of NA of an ideal lens with unity transmission efficiency in the air is shown in Fig. 2F (*32*). Considering isotropic SPEs having the same radiation in all directions (i.e., uniform radiation), a high-NA lens of NA=0.869, corresponding to the maximum collection angle of 120º, leads to an ideal collection efficiency of 0.3. To experimentally verify the high NA of our metalens, we measure the point spread function (PSF) of the metalens based on an incident collimated beam (Supplementary Note 3). The cyan star marker in the figure denotes the fabricated metalens (NA of 0.869) collection efficiency, which considers both the average transmission efficiency of meta-atoms (~0.94) and the substrate transmittance (~0.91). Fig. 2F inset shows the resulting PSF line profiles of the metalens in both experiment (scatter line) and theoretical calculation (solid line), showing a great agreement verifying that our metalens has a high NA of 0.869. The small side lobes in the characterised PSF profiles are mainly due to the off-axis aberration in a tight focus (*33*).

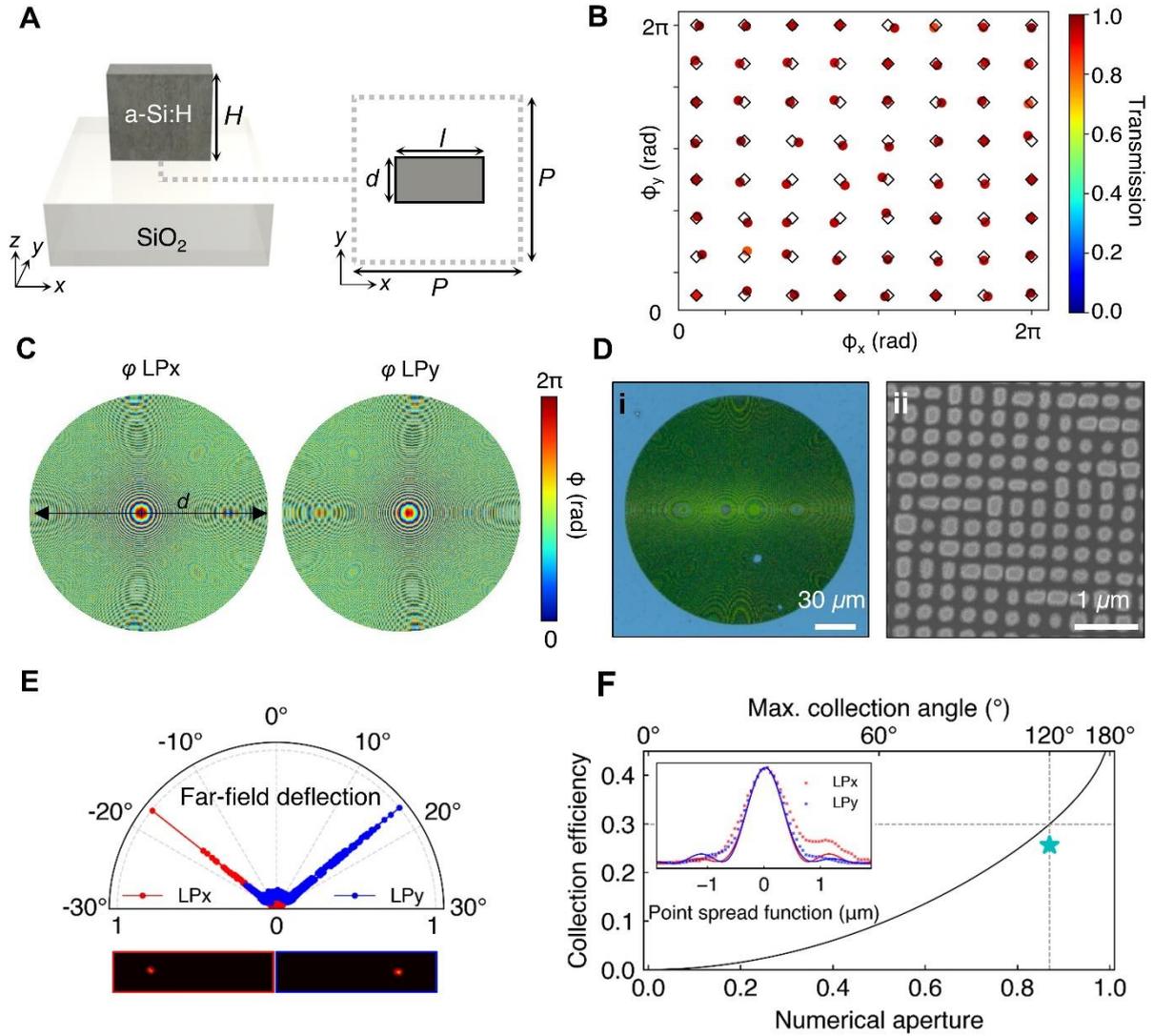

**Fig. 2. Design and characterisation of a polarisation beam-splitting metalens.** (**A**) Schematic of the meta-atom blocks from low-loss a-Si:H on silica. (**B**) Calculated phase delay and transmission library of meta-atoms covering $2\pi$ for both LPx and LPy light. The average transmission is 94%. (**C**) Implemented phase maps for the metalens and gradient functions for LPx and LPy incidence. (**D**) Images of the fabricated metalens taken with an (i) optical microscope and (ii) scanning electron microscope. (**E**) Polar plot of the measured far-field deflection of the metalens under LPx and LPy dipole-like illumination. The bottom images are associated back-focal-plane image patterns. (**F**) Calculated collection efficiency of an isotropic emitter in free space as a function of numerical aperture of the collecting lens. A cyan star marks our metalens collection efficiency with NA=0.869. Inset: point spread function of the metalens under the LPx and LPy plane-wave incidence, where scatter and solid lines represent the experimental and theoretical characterisations, respectively.

## 2.2 Characterisation of quantum emission with a polarisation beam-splitting metalens

To prepare the hBN SPEs, the hBN solution containing SPEs is prepared on a glass substrate (see Methods). We choose SPEs with emission at ~ 610 nm, to match the designed metalens. First, the emitters are characterised using a conventional confocal microscope with a Hanbury Brown-Twiss interferometer to confirm the quantum nature of emitted light, as shown in Fig. 3A. Briefly, the hBN emitters are excited using a green (532 nm) laser and the emission is collected using same objective. The chosen SPE is shown in Fig. 3B, with a dominant zero-phonon line (ZPL) centred at 591 nm and a minor phonon sideband (PSB). Photon anti-bunching behaviour is confirmed by recording the second-order correlation function, $g^{(2)}(\tau)$, with a background corrected $g^{(2)}(0) = 0.05$. Finally, the SPE dipole orientation is confirmed to be 106° (black arrow) with respect to the laboratory frame, as confirmed by rotating a linear polariser in the collection path (Fig. 3C).

We now utilise the metalens for multidimensional structuring of the quantum emission of the precharacterised hBN SPE. First, the metalens was used as a collection lens in the transmission path and the LPx direction is set to horizontal (90°) as shown in Fig. 3C (red arrow). An SPE dipole-to-metalens LPx intersection angle of 16° is hence derived. As expected, the SPE emission passes through the polarisation beam-splitting metalens and splits into LPx (red curve) and LPy (blue curve) as shown in Fig. 3D. Both photon streams result in sharp spectra with peaks centred at 591 nm, corresponding to the ZPL of the SPE. LPx and LPy polarisation are confirmed with a polariser and have been used to calibrate the metalens mounting angle. Note, the peak area ratio is $A_{LPy}/A_{LPx} = 0.19$ which is slightly higher than the anticipated dipole projection ratio $(\tan(16°))^2 = 0.08$ in the orthogonal directions. One possible explanation is the variation of free space to fibre collection efficiency at each channel, as will be later confirmed by back-focal plane imaging. The inset in Fig. 3D shows confocal photoluminescence (PL) scans of the SPE as imaged through the metalens in both LPx and LPy channels, normalised to the LPx maximum. The distinct emission from SPE as well as their signal-to-noise ratio are clearly visualised for each polarisation channel. The missing PSB peak is caused by the wavelength selectivity of the metalens, more discussion can be found in Supplementary Note 4.

To clarify that the detected photons are from the same SPE, 2nd order cross-correlation measurements between these PL channels are performed. Fig. 3 (E and F) show cross-correlation measurements between the emitter PL in reflection (employing a traditional objective) versus LPx and LPy (employing the metalens as the collecting lens). Fig. 3G shows the cross-correlation between the two polarised channels LPx and LPy (using the metalens to collect the light). Note that the correlation between LPx and LPy channels illustrates the single-photon emission of the source. The same SPE with its corresponding quantum behaviour is verified across all three measurements, as all the $g^{(2)}(0)$ values approximately equal to 0.1 (Supplementary Note 5). These successful measurements confirm the potential of metalens for low-intensity light collection - as demonstrated with quantum emitters in hBN. Furthermore, the independent sample substrate enables its use as a confocal mapping lens. To this extent, the results of other SPEs are shown in Supplementary Note 6.

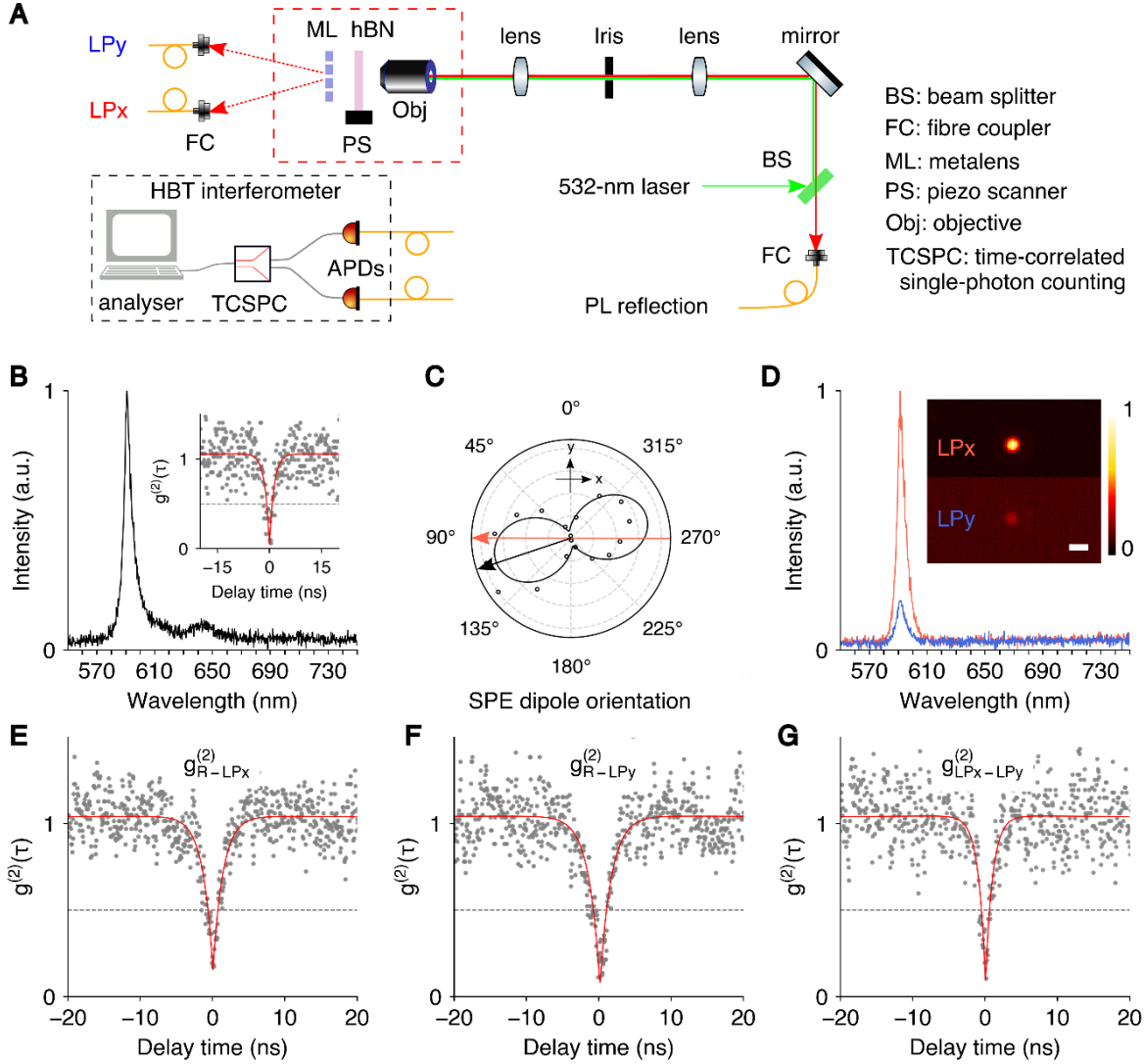

**Fig. 3. Characterisation of hBN SPEs with the polarisation beam-splitting metalens.** (**A**) A schematic optical setup for the photonic measurement. (**B**) PL collected via reflection channel (traditional objective) with a ZPL at ~ 591 nm. Inset, $g^{(2)}(\tau)$ confirmed the quantum nature of the SPE. (**C**) emitter dipole orientation, measured from the reflection channel where 0° is the y (vertical) direction. The metalens orientation is set to 90° (red arrow). The emitter dipole orientation (black arrow) of 106° is derived after fitting with a sine function. (**D**) PL spectra collected from LPx (red curve) and LPy (blue curve) polarised channels employing the metalens. Insets are corresponding PL raster maps of the emitter. Colour bars in two maps are set to the same for a clear intensity contrast. Scale bars are 1 μm. (**E-G**) are cross correlation antibunching measurements from different channels. (**E**) reflection PL and LPx correlation. (**F**) Reflection PL and LPy correlation. (**G**) LPx and LPy correlation. $g^{(2)}(\tau)$ data is background corrected.

## 2.3 Arbitrarily structuring quantum emission

After confirming the beam-splitting metalens efficiently collects and spatially separates polarisation streams of the emitted light from SPEs, we imprint different functionalities on the metalenses for multi-dimensional structuring of quantum emission. We design and fabricate four different metalenses with distinct functionalities and increasing complexity, as shown schematically in Fig. 4A. By employing the 8-step (16-step) phase meta-atoms, Metalens 1 (2) deflects LPx and LPy to angles of ±12.58° (±6.29°) along the x-direction, respectively. Metalens 3 deflects LPx and LPy in arbitrary directions, shifted from the centre, and finally, Metalens 4 encodes different helical phase fronts carrying distinctive OAM values ($l$=3 and $l$=-1, in our case) - in addition to the directionality and polarisation. Fig. 4B shows the far-field optical images of the fabricated four metalenses and their characteristic interference patterns. We then employ these metalenses to image the same SPE, as presented in Fig. 3.

The results of the four metalenses are shown in Fig. 4C. For convenience and clarity, we rotate the emitter to 146°, resulting in a dipole-to-metalens intersection angle of 56°, to balance the emission rate of the two channels (Supplementary Note 7). The white dash circles in the images represent the maximum splitting angle of 25°, while the inset red (blue) circle is a zoomed-in image of the LPx (LPy) spot. Using Metalens 1 we observed two bright spots, and the PL signals (summed in Ky direction) gave a clear peak profile of the two spots with an intensity ratio of $A_{LPy}/A_{LPx} = 1.8$, indicating a dipole projection angle of 53° which aligns well with the measured 56°. Metalens 2 demonstrates axial symmetric deflection with a half-splitting angle compared to Metalens 1. As expected, the emitters are imaged closer, with a smaller deflection angle and spot intensity ratio of $A_{LPy}/A_{LPx} = 1.8$. Metalens 3 splits the incoming SPE asymmetrically, with two spots falling into the same side and also deflected in the y axis ($A_{LPy}/A_{LPx} = 13.7$). The combination of Metalenses 1-3 hence allows for complete direction and polarisation control of the imaged SPEs. As a final demonstration, we employ Metalens 4 to show two doughnut beams (zoomed-in in the insets) that correspond to the OAM with topological charges of $l = 3$ and $l = -1$ for the LPx and LPy channels, respectively. Since the larger OAM leads to larger doughnut intensity distribution in the momentum space, LPy becomes brighter than LPx ($A_{LPy}/A_{LPx} = 2.0$). Furthermore, we experimentally confirm that the two structured beams carry the correct topological charges (Supplementary Note 8). Moreover, structuring SPE with the OAM property is achieved from the phase manipulation, which conserves the sub-Poissonian photon statistics and does not affect the antibunching nature of the photons. As a result, our metalens conclusively demonstrates a complete control of single photons in directionality, polarisation and OAM dimensions, simultaneously.

To support these results, we have also performed scalar wave propagation calculations. Complex-amplitude profiles were extracted from the four metalens with two polarisation channels, then modified by assuming the dipole source is located at the exact focal position (subtraction of lens phase in each profile). First, each modified profile with the two polarisation channels was used to evaluate far-field profiles using Rayleigh-Sommerfeld (RSD) diffraction integral. Then, far-field images in the LPx and LPy channel were directly integrated by considering the polarisation angle of the dipole source (Fig. 4C bottom panel). Each metalens performs the desired function well, showing a clear collimated beam that is successfully structured as designed. Compared to previous demonstrations (*23, 24, 25*), this work presents for the first time multi-dimensional shaping of quantum emission by using a single metalens structure, capable of simultaneous control of the linear momentum, polarisation, and OAM of SPEs (Supplementary Table S1).

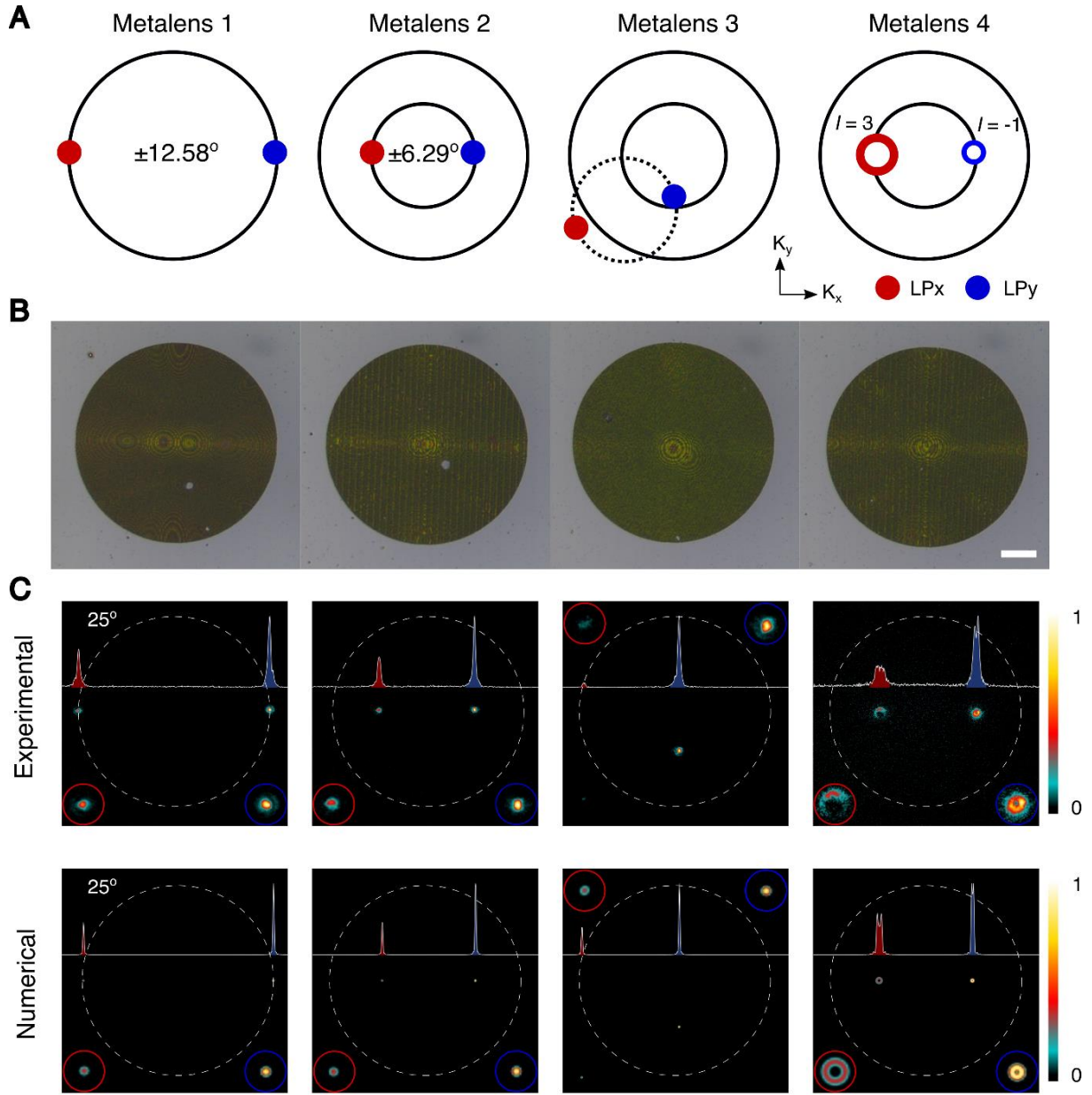

**Fig. 4. Arbitrary structured quantum emission with polarising beam-splitting metalenses.** (**A**) Schematics of different metalens functionalities. The circles represent desired positions for orthogonal linear polarisations in the k space. The doughnut-shaped spots indicate the generation of OAM. (**B**) Far-field optical images of the fabricated metalens. Scale bar is 30 μm. (**C**) Experimental (top) and numerically calculated (bottom) results of the SPE far-field back focal imaging. Curves in c represent the vertically accumulated PL intensity of the LPx (filled with red) and LPy (filled with blue). The red and blue circles are zoomed-in insets for highlighting the intensity distributions of the orthogonal polarisations.

## 3 Discussion

In this work, we proposed and implemented a multifunctional metalens to arbitrarily structure quantum light sources at room temperature. Realised with bright and linearly polarised SPEs in hBN, our designed and fabricated polarisation beam-splitting metalens allows independent wavefront shaping of orthogonal linear polarisations. We further experimentally demonstrated arbitrary directionality control of single photon emission, as well as imprinting different OAM modes onto its orthogonal polarisations. As such, our demonstration offers a new platform to use ultrathin meta-optics for arbitrary wavefront shaping of quantum emission in multiple degrees of freedom at room temperature. It may provide new insights into the field of quantum information science - specifically manipulating photon polarisations has great impact for quantum cryptography and entanglement distribution with improved filtering. The polarisation separation is vital, for future employment of hBN SPEs for polarisation entangled photon pair generation, as it is commonly used for down converted photon sources(*34*). Notably, a major reason of choosing SPE instead of other types of point-like sources is because manipulating quantum source instead of classic photoluminescence meets the needs of quantum photonics. For future quantum applications, we will need not only abundant quantum sources with various on-demand properties but also miniaturized device footprint for good scalability. To the best of our knowledge, this work has achieved the highest degrees of freedom in the manipulation of SPEs in the air environment at a room temperature.

Although the focus here is the multi-functionality and flexibility of the metalens to image a selected SPE, integration of hBN SPE to metalens could also be possible by bulk engineering or addition of a transparent spacer(*35*), adapting device architectures and aligning approaches developed earlier(*26*). Photon collection of solid-state SPEs is an outstanding challenge which may hinder practical applications. Indeed, various strategies to enhance photon collection including the use of distributed Bragg reflectors(*36*), perturbed photonic crystal cavities(*37*), circular grating cavities(*38, 39*), nano-antennas(*40*), and solid immersion lenses(*25*) etc. We believe their implementation with our demonstrated wavefront-shaping metasurface is not a trivial task and beyond the scope of current work. In the current work, we focus on improving functionality, rather than focusing solely on collection enhancement. We note, that our proposed hybrid hBN – silica metalens can be implemented together with any additional photon collector.

Moreover, our confocal mapping metalens endows a direct access to the OAM states of single photons, offering a compact platform for high-dimensional quantum entanglement to increase the quantum information capacity(*41-44*). Future extension of the metalens to realise higher-order structured vector beams(*41, 45, 46*) could enable the generation of high-dimensional single-photon hybrid quantum states. Furthermore, future integration of structured SPE sources with a reliable transmission environment such as optical fibres(*47*) could promise a quantum network with higher information capacity, robustness to noise and better security. Lastly, we envision that combining the spin defects in hBN with the proposed multifunctional metalens may enable efficient spin photon readout and remote quantum sensing(*48-52*).

# 4 Methods

## 4.1 Numerical calculations

The propagation phase delay and field profiles of the meta-atoms were calculated using an in-house built RCWA solver. The far-field propagation of the four metalenses was calculated using scalar diffraction, specifically the RSD integral method. The input amplitude and phase profiles for each metalens were retrieved from the S parameters of the spatially arranged meta-atoms.

## 4.2 Sample preparation

*Metalens fabrication.* Plasma-enhanced chemical vapour deposition (PECVD, BMR Technology HiDEP-SC) was used to deposit a 550-nm-thick hydrogenated amorphous silicon (a-Si:H) layer on a 500-μm thick silica wafer. A gas flow rate, chamber pressure and operating temperature were set to be 10 sccm for $SiH_4$ and 75 sccm for $H_2$, 25 mTorr, and 200 °C, respectively. The standard electron beam lithography process (ELIONIX, ELS-7800) was used to transfer the designed metalens onto the positive photoresist (495 PMMA A6, Micro-Chem). An acceleration voltage and beam current were set to be 80 kV and 500 pA, respectively. The exposed photoresist patterns were developed in methyl isobutyl ketone/IPA (1:3) solution for 10 min at 0 °C. The electron-beam evaporation (KVT, KVE-KVE-ENS4004) was used to deposit a 60-nm-thick chromium (Cr) mask and residual unexposed photoresists were stripped using acetone. The Cr mask pattern was transferred onto the a-Si:H layer using a dry etching (DRM85DD, TEL). The residual Cr mask was removed by Cr etchant (CR-7).

*hBN sample preparation.* hBN nanoflake solution (few layers with submicrometer lateral size, Graphene Supermarket) was drop casted on a gold-marked thin glass coverslip, and then dried on hotplate at 50°C for 5 min. Following by 400°C annealing on hotplate for 3 hrs and 5 min UV ozone plasma to oxidise the potential organic residuals. SPEs within the target spectral range (610±20 nm) were then pre checked by a lab-built confocal microscope and later relocated with the assistance of gold markers. For the specific sample used in this work, we can roughly find less than ten emitters per scan (50x50 $\mu m^2$).

## 4.3 Experimental setup

*Metalens characterisation.* The schematics of optical setups for metalens characterisation are illustrated in Supplementary Note 3. A continuous-wave laser beam at 633 nm (CPS635, Thorlabs) was first passed through a spatial filter system which consists of two concave lenses and a 5 *μ*m pinhole (P5CB, Thorlabs). Then the beam was linearly polarised using a linear polariser (LPVISE100-A, Thorlabs), and the polarisation angle is tuned using a halfwave plate (AHWP10M-600, Thorlabs). For PSF imaging, a weakly focused beam using a 4X/0.13 NA objective (LMPlanFL N, Olympus) was incident on the metalens. The focused beam by the metalens was captured using an imaging setup comprised of a 100x/0.8 NA objective (LMPlanFL N, Olympus), tube lens (TTL180-A, Thorlabs), and CCD (INFINITY2-1RC, Infinity). For imaging far-field patterns, the beam was tightly focused by the 100X/0.8 NA objective to imitate dipole emission. The dipole-like beam was then collected and collimated by the metalens. A back focal image of the collimated beam was formed at the back focal plane using a 20x/0.45 NA objective (LMPlanFL N, Olympus), and captured using two concave lenses and sCMOS camera (Panda 4.2 bi UV, PCO).

*SPE-metalens PL characterisation.* The schematics of a lab-built optical setup for SPE-metalens characterization were illustrated in Fig. 3A**.** A 532-nm CW laser (Gem 532TM, Laser Quantum Ltd.) was used to pump the SPE through a 100x/0.9NA air objective (TU Plan Fluor, Nikon), reflected PL signal was filtered by a beam splitter and long pass filter before coupled into a 62.5-μm core multimode fibre and then connected to a spectrometer (Princeton Instruments, Inc.). Or be split by a 50:50 fibre splitter and went into two avalanche photodiodes (APDs, SPCM-AQRH-16, Excelitas Technologies) for Hanbury Brown and Twiss measurements. The transmission PL was collected by the metalens. Given the short focal distance of the metalens, the hBN side and metalens side were placed face to face. Next, the metalens location was first confirmed by the supercontinuum white light laser (SuperK Fianium, NKT) with the wavelength being set to the same wavelength of the emitter. Two collimated beam spots can be observed once the metalens focal plane reaches the focal spot of the incoming laser. The sample mounting and aligning platform are shown in Supplementary Note 7. The two beams were then independently coupled into two multimode fibres (core size 62.5 μm) and then can be connected either to a spectrometer or an APD. Photon correlation was done by a time-correlated single-photon counting module (Picoharp 300, PicoQuant). All measurements were done under an excitation power of 500 μW (before objective). And 568-nm long pass filters are used to filter out lasers in all collections.

*SPE-metalens back focal imaging characterisation.* Back-focal plane imaging allows the direct measurement of the far-field intensity distribution emanating from the single-photon source. The optical setup used is similar with the metalens characterisation setup (Supplementary Note 3), where a back focal image of the collimated beam is formed at the back focal plane using a 20x/0.4 NA objective (LMPlanFL N, Olympus). Then acquired by an EMCCD (iXon Ultra 888) placed at a distance of 4$f$ away from the objective on a path, separate from the main confocal collection pathway.

## Declarations

### Data availability

The main data supporting the findings of this study are available within the article and its Supplementary Information files. Extra data are available from the corresponding author upon reasonable request.

### Competing interests

The authors declare that they have no competing interests.

### Funding

This work was supported by Australian Research Council (CE200100010, DE220101085, DP220102152) and the Office of Naval Research Global (N62909-22-1-2028)(I.A.).

S.A.M. acknowledges the Lee-Lucas Chair in Physics.

J.R. acknowledges the POSCO-POSTECH-RIST Convergence Research Center program funded by POSCO, and the National Research Foundation (NRF) grants (NRF-2022M3C1A3081312, NRF-2022M3H4A1A02085335, NRF-2022M3H4A1A02074314, NRF-2022K1A3A1A25081970, NRF-2019R1A2C3003129, CAMM-2019M3A6B3030637, NRF-2019R1A5A8080290) funded by the Ministry of Science and ICT of the Korean government. J.H.K. acknowledges the POSTECH Alchemist fellowship.

### Author Contributions

I.A., H.R., and J.R conceived the concept of this project. C.L. prepared SPEs and conducted the SPE-metalens measurements. J. J. conducted optical characterisation of the metalenses. T.B. and J.J. designed the metalenses. J.H.K. and J.K.K. fabricated the metalenses. C.L., T.Y. and M.N. built the photonic characterisation platform for the SPE-metalens system. C.L., J.J., T.B. and H.R. analysed the data. All authors discussed the results and wrote the manuscript.

## Acknowledgements

The authors thank Milos Toth, Yongliang Chen and Simon White for their fruitful discussion.

## Corresponding authors

Correspondence to Chi Li: chi.li@uts.edu.au; Junsuk Rho: jsrho@postech.ac.kr and Haoran Ren: haoran.ren@monash.edu

## References

1. D. D. Awschalom, R. Hanson, J. Wrachtrup, B. B. Zhou, Quantum technologies with optically interfaced solid-state spins. *Nature Photonics* **12**, 516-527 (2018).
2. X. Liu, M. C. Hersam, 2D materials for quantum information science. *Nature Reviews Materials* **4**, 669-684 (2019).
3. I. Aharonovich, J. P. Tetienne, M. Toth, Quantum Emitters in Hexagonal Boron Nitride. *Nano Lett* **22**, 9227-9235 (2022).
4. J. D. Caldwell *et al.*, Photonics with hexagonal boron nitride. *Nature Reviews Materials* **4**, 552-567 (2019).
5. F. Hayee *et al.*, Revealing multiple classes of stable quantum emitters in hexagonal boron nitride with correlated optical and electron microscopy. *Nat Mater* **19**, 534-539 (2020).
6. P. Senellart, G. Solomon, A. White, High-performance semiconductor quantum-dot single-photon sources. *Nat Nanotechnol* **12**, 1026-1039 (2017).
7. A. H. Dorrah, F. Capasso, Tunable structured light with flat optics. *Science* **376**, eabi6860 (2022).
8. M. L. Tseng *et al.*, Vacuum ultraviolet nonlinear metalens. *Sci Adv* **8**, eabn5644 (2022).
9. T. Santiago-Cruz *et al.*, Resonant metasurfaces for generating complex quantum states. *Science* **377**, 991-995 (2022).
10. A. S. Solntsev, G. S. Agarwal, Y. S. Kivshar, Metasurfaces for quantum photonics. *Nature Photonics* **15**, 327-336 (2021).
11. J. Ni *et al.*, Multidimensional phase singularities in nanophotonics. *Science* **374**, eabj0039 (2021).
12. L. Li *et al.*, Metalens-array-based high-dimensional and multiphoton quantum source. *Science* **368**, 1487-1490 (2020).
13. K. Wang *et al.*, Quantum metasurface for multiphoton interference and state reconstruction. *Science* **361**, 1104-1108 (2018).
14. H. Ren *et al.*, An achromatic metafiber for focusing and imaging across the entire telecommunication range. *Nat Commun* **13**, 4183 (2022).
15. T. Badloe, I. Kim, Y. Kim, J. Kim, J. Rho, Electrically Tunable Bifocal Metalens with Diffraction-Limited Focusing and Imaging at Visible Wavelengths. *Adv Sci* **8**, e2102646 (2021).
16. W. J. Joo *et al.*, Metasurface-driven OLED displays beyond 10,000 pixels per inch. *Science* **370**, 459-463 (2020).
17. X. Ni, A. V. Kildishev, V. M. Shalaev, Metasurface holograms for visible light. *Nature Communications* **4**, 1-6 (2013).
18. H. Ren *et al.*, Complex-amplitude metasurface-based orbital angular momentum holography in momentum space. *Nat Nanotechnol* **15**, 948-955 (2020).

19. L. Jin *et al.*, Dielectric multi-momentum meta-transformer in the visible. *Nat Commun* **10**, 4789 (2019).
20. X. Fang, H. Ren, M. Gu, Orbital angular momentum holography for high-security encryption. *Nature Photonics* **14**, 102-108 (2020).
21. S. So *et al.*, Multicolor and 3D Holography Generated by Inverse-Designed Single-Cell Metasurfaces. *Advanced Materials*, 2208520 (2022).
22. S. Q. Li *et al.*, Phase-only transmissive spatial light modulator based on tunable dielectric metasurface. *Science* **364**, 1087-1090 (2019).
23. I. Kim *et al.*, Holographic metasurface gas sensors for instantaneous visual alarms. *Sci Adv* **7**, eabe9943 (2021).
24. F. Yang, P. K. Jha, H. Akbari, H. C. Bauser, H. A. Atwater, A hybrid coupler for directing quantum light emission with high radiative Purcell enhancement to a dielectric metasurface lens. *Journal of Applied Physics* **130**, 163103 (2021).
25. T. Y. Huang *et al.*, A monolithic immersion metalens for imaging solid-state quantum emitters. *Nat Commun* **10**, 2392 (2019).
26. Y. Bao *et al.*, On-demand spin-state manipulation of single-photon emission from quantum dot integrated with metasurface. *Sci Adv* **6**, eaba8761 (2020).
27. C. Wu *et al.*, Room-temperature on-chip orbital angular momentum single-photon sources. *Sci Adv* **8**, eabk3075 (2022).
28. E. N. Knall *et al.*, Efficient Source of Shaped Single Photons Based on an Integrated Diamond Nanophotonic System. *Phys Rev Lett* **129**, 053603 (2022).
29. B. Chen *et al.*, Bright solid-state sources for single photons with orbital angular momentum. *Nat Nanotechnol* **16**, 302-307 (2021).
30. Y. Yang *et al.*, Revealing Structural Disorder in Hydrogenated Amorphous Silicon for a Low-Loss Photonic Platform at Visible Frequencies. *Adv Mater* **33**, e2005893 (2021).
31. G. Yoon, K. Kim, D. Huh, H. Lee, J. Rho, Single-step manufacturing of hierarchical dielectric metalens in the visible. *Nat Commun* **11**, 2268 (2020).
32. W. Barnes *et al.*, Solid-state single photon sources: light collection strategies. *The European Physical Journal D-Atomic, Molecular, Optical and Plasma Physics* **18**, 197-210 (2002).
33. Y. Guo *et al.*, High-efficiency and wide-angle beam steering based on catenary optical fields in ultrathin metalens. *Advanced Optical Materials* **6**, 1800592 (2018).
34. P. G. Kwiat *et al.*, New high-intensity source of polarization-entangled photon pairs. *Phys Rev Lett* **75**, 4337-4341 (1995).
35. J. L. O'brien, A. Furusawa, J. Vučković, Photonic quantum technologies. *Nature Photonics* **3**, 687-695 (2009).
36. F. Basso Basset *et al.*, Quantum key distribution with entangled photons generated on demand by a quantum dot. *Sci Adv* **7**, eabe6379 (2021).
37. M. Toishi, D. Englund, A. Faraon, J. Vuckovic, High-brightness single photon source from a quantum dot in a directional-emission nanocavity. *Opt Express* **17**, 14618-14626 (2009).
38. J. E. Froch *et al.*, Coupling Spin Defects in Hexagonal Boron Nitride to Monolithic Bullseye Cavities. *Nano Lett* **21**, 6549-6555 (2021).
39. J. Liu *et al.*, A solid-state source of strongly entangled photon pairs with high brightness and indistinguishability. *Nat Nanotechnol* **14**, 586-593 (2019).
40. X.-L. Chu *et al.*, Experimental realization of an optical antenna designed for collecting 99% of photons from a quantum emitter. *Optica* **1**, 203-208 (2014).
41. A. Mair, A. Vaziri, G. Weihs, A. Zeilinger, Entanglement of the orbital angular momentum states of photons. *Nature* **412**, 313-316 (2001).
42. R. Fickler *et al.*, Quantum entanglement of high angular momenta. *Science* **338**, 640-643 (2012).
43. M. Malik *et al.*, Multi-photon entanglement in high dimensions. *Nature Photonics* **10**, 248-252 (2016).

44. T. Badloe, S. Lee, J. Rho, Computation at the speed of light: metamaterials for all-optical calculations and neural networks. *Advanced Photonics* **4**, 064002 (2022).
45. A. Forbes, M. de Oliveira, M. R. Dennis, Structured light. *Nature Photonics* **15**, 253-262 (2021).
46. T. Stav *et al.*, Quantum entanglement of the spin and orbital angular momentum of photons using metamaterials. *Science* **361**, 1101-1104 (2018).
47. D. Cozzolino *et al.*, Orbital Angular Momentum States Enabling Fiber-based High-dimensional Quantum Communication. *Physical Review Applied* **11**, 064058 (2019).
48. A. Gottscholl *et al.*, Spin defects in hBN as promising temperature, pressure and magnetic field quantum sensors. *Nat Commun* **12**, 4480 (2021).
49. H. L. Stern *et al.*, Room-temperature optically detected magnetic resonance of single defects in hexagonal boron nitride. *Nat Commun* **13**, 618 (2022).
50. N. Chejanovsky *et al.*, Single-spin resonance in a van der Waals embedded paramagnetic defect. *Nat Mater* **20**, 1079-1084 (2021).
51. X. Gao *et al.*, Nuclear spin polarization and control in hexagonal boron nitride. *Nat Mater* **21**, 1024-1028 (2022).
52. I. O. Robertson *et al.*, Imaging current control of magnetization in $Fe_3GeTe_2$ with a widefield nitrogen-vacancy microscope. *2D Materials* **10**, 015023 (2022).